\documentclass[aps,twocolumn]{revtex4}

\RequirePackage[T2A]{fontenc}

\usepackage{amssymb}
\usepackage{amsmath}
\usepackage{bm}
\usepackage[dvips]{epsfig}
\usepackage{graphicx}
\usepackage{textcomp}

\begin{document}

\title{On the level lines of bilayer potentials with a finite 
number of harmonics} 

\author{A.Ya. Maltsev}

\affiliation{
\centerline{\it L.D. Landau Institute for Theoretical Physics}
\centerline{\it 142432 Chernogolovka, pr. Ak. Semenova 1A,
maltsev@itp.ac.ru}}

\begin{abstract}
 We consider ``bilayer'' potentials formed by the superposition of two 
periodic potentials in a plane. We assume that both potentials share 
the same rotational symmetry and employ an approximation involving a 
finite number of harmonics in their Fourier expansions. An analysis of 
the level lines of the resulting quasiperiodic potential (a problem 
posed by S.P. Novikov) near the percolation threshold makes it possible 
to describe the behavior of large-scale closed level lines in this limit. 
The results presented here can also be applied to broader classes of 
bilayer potentials that possess ``good'' analytical properties. 
\end{abstract}

\maketitle

\section{Introduction}

 In this work, we investigate two-dimensional potentials obtained by 
superimposing two periodic potentials, $\, V_{1} (x, y) \, $ and 
$\, V_{2} (x, y) \, $, which share the same rotational symmetry (of order three, 
four, or six). More precisely, we will consider families of potentials
$$V \left( x, y, \, \alpha , {\bf a} \right) \,\,\, = \,\,\, 
V_{1} ( x, y ) \,\, + \,\, V_{2} \left( x, y, \, \alpha, {\bf a} \right) 
\,\,\, , $$
where the potential
$$ V_{2} \left( {\bf r}, \, \alpha , {\bf a} \right) \,\,\, = \,\,\, 
U \left( \pi_{-\alpha} ({\bf r} - {\bf a}) \right) $$
is derived from a fixed potential $\, U ({\bf r}) \, $ by applying a rotation 
through an angle $\, \alpha \, $ (about the origin) and a translation by a vector 
$\, {\bf a} \, $ in the plane $\, \mathbb{R}^{2} \, $ (here and throughout the 
following, the operator $\, \pi_{\alpha} \, $ denotes a rotation of an object by 
an angle $\, \alpha \, $ about the origin). As previously mentioned, we assume that 
the (periodic) potentials $\, V_{1} (x, y) \, $ and $\, U (x, y) \, $ possess the 
same rotational symmetry; however, their periods are not required to be equal.

 In the generic case (in particular, when the periods of $\, V_{1} (x, y) \, $
and $\, U (x, y) \, $ are incommensurable), the potential 
$\, V \left( {\bf r} , \, \alpha , {\bf a} \right) \, $
is a quasiperiodic function with 4 quasiperiods. Here, we define a function 
$\, f (x, y) \, $ on $\, \mathbb{R}^{2} \, $ as a quasiperiodic function with 
$\, N \, $ quasiperiods if it is the restriction of some $\, N$ - periodic function 
$\, F \left( X^{1} , \dots , X^{N} \right) \, $ defined in $\, \mathbb{R}^{N} \, $
to a subspace $\, \mathbb{R}^{2} \,$ via an affine embedding
$\, \mathbb{R}^{2} \subset \mathbb{R}^{N} \,$:
$$f ({\bf r}) \,\,\, = \,\,\, F \left( {\bf X} \right) 
\big|_{\mathbb{R}^{2} \subset \mathbb{R}^{N}} $$

 For certain values of $\, {\bf a} \, $ (specifically, when some of the centers 
of symmetry of the potentials $\, V_{1} ({\bf r}) \, $ and $\, V_{2} ({\bf r}) \, $
coincide), the potentials $\, V \left( {\bf r} , \, \alpha , {\bf a} \right) \, $
exhibit exact rotational symmetry. The corresponding values of $\, {\bf a} \, $ 
form a dense subset of $\, \mathbb{R}^{2} \, $;
however, for a generic choice of $\, {\bf a} \, $, the potentials 
$\, V \left( {\bf r} , \, \alpha , {\bf a} \right) \, $
do not possess exact rotational symmetry. In standard terminology 
(see, e.g., \cite{LePiunSad}), the family 
$\, V \left( {\bf r} , \, \alpha , {\bf a} \right) \, $
is said to possess quasicrystalline rotational symmetry (of the corresponding order).

 In some cases, the potentials 
$\, V \left( {\bf r} , \, \alpha , {\bf a} \right) \, $ can be periodic. This occurs 
when the periods of the potentials $\, V_{1} ({\bf r}) \, $ and $\, U ({\bf r}) \, $
are commensurable, and the angles $\, \alpha \, $ take on specific (``magic'') values. 
In the generic case, the potentials 
$\, V \left( {\bf r} , \, \alpha , {\bf a} \right) \, $
are not periodic; however, they can often be approximated by periodic potentials 
(with large periods).

 Here, we will examine a problem posed by S.P. Novikov: the description of the 
(global) geometry of the level sets
\begin{equation}
\label{LevelSet}
f (x, y) \,\,\, = \,\,\, {\rm const} 
\end{equation}
for generic (non-periodic) potentials 
$\, V \left( {\bf r} , \, \alpha , {\bf a} \right) \, $.
The most challenging aspect of Novikov’s problem is the description of the open 
(unbounded) connected components of the sets defined by (\ref{LevelSet}). At the 
same time, however, the description of the bounded components of (\ref{LevelSet}) - 
particularly those that are large in size and adjacent to the unbounded components - 
also plays a significant role in the overall study of this problem.

\vspace{1mm}

 The most extensively studied case of the Novikov problem is that of functions 
with 3 quasiperiods (\cite{MultValAnMorseTheory,zorich1,dynn1992,Tsarev,
dynn1,zorich2,DynnBuDA,dynn2,dynn3,PismaZhETF,ZhETF1997,UFN,BullBrazMathSoc,
StatPhys,DeLeoPhysLettA,DeLeoPhysB,TrMIAN,DeLeo2017,zorich3,DeLeo1,DeLeo2,
DeLeo3,DeLeoDynnikov1,Dynnikov2008,DeLeoDynnikov2,Skripchenko1,Skripchenko2,
DynnSkrip1,DynnSkrip2,AvilaHubSkrip1,AvilaHubSkrip2,DynMalNovUMN,DynHubSkrip}),
which plays an important role in the theory of galvanomagnetic phenomena in metals. 
The general case of functions with 4 quasiperiods was investigated in the works 
\cite{NovKvazFunc,DynNov}, which contain the most profound results for $\, N = 4 \, $. 
A number of general results for the case of an arbitrary $\, N \, $ were also obtained 
in \cite{DynMalNovUMN,BigQuas,Sirius}. 

\vspace{1mm}

 The case of the superposition of two periodic potentials in a plane corresponds 
to a specific instance of $\, N = 4 \, $. This case plays an important role in the 
modern physics of two-dimensional systems (in particular, the physics of ``bilayer'' 
systems) and has been investigated in the works \cite{AnnPhys,Superpos,SymSuperPos}. 

\vspace{1mm}

 A distinctive feature of (non-periodic) potentials 
$ V \left( {\bf r} , \, \alpha , {\bf a} \right) \, $ that possess (quasicrystalline) 
rotational symmetry is that unbounded components of the set
\begin{equation}
\label{VcLevel}
V \left( {\bf r} , \, \alpha , {\bf a} \right) \,\,\, = \,\,\, c 
\end{equation}
(or bounded components of arbitrarily large size) can occur here only at a single 
specific level, $\, c = c_{0} \, $ (\cite{Superpos}).

 For given potentials $\, V_{1} ({\bf r}) \, $ and $\, U ({\bf r}) \, $, the value 
$\, c_{0} \, $ is independent of the value of $\, {\bf a} \, $; thus, we have 
$\, c_{0} = c_{0} (\alpha) \, $. For $\, c \neq c_{0} \, $, the sizes of all connected 
components of (\ref{VsLevel}) are bounded (for all values of ${\bf a}$) by a constant
$\, D = D (\alpha , c) \, $, which may tend to infinity as 
$\, c \rightarrow c_{0} (\alpha) \, $. 

 The open level lines
$$ V \left( {\bf r} , \, \alpha , {\bf a} \right) \,\,\, = \,\,\,
c_{0} (\alpha) $$
have a highly complex shape, resembling a random walk in the plane 
$\, \mathbb{R}^{2} \, $. Such properties of the potentials 
$ V \left( {\bf r} , \, \alpha , {\bf a} \right) \, $ make them similar to random 
potentials in the plane, which exhibit analogous behavior of their level lines.

\vspace{1mm}

 A more detailed comparison of the level sets (\ref{VcLevel}) with the level 
sets of random potentials, however, reveals certain differences between these 
two classes of functions. One such difference manifests, in particular, in the 
rate at which the sizes of the bounded components of (\ref{VcLevel}) - and the 
bounded level sets of random potentials - grow as $\, c \, $ approaches the 
value $\, c_{0} \, $. 

 In \cite{SymSuperPos}, such an estimate was derived for potentials 
$\, V \left( {\bf r} , \, \alpha , {\bf a} \right) \, $ that admit an (additional) 
reflection symmetry. In this case, for almost all angles $\, \alpha \, $, the growth 
rate of the sizes of the closed level components (\ref{VcLevel}) can be estimated 
by the asymptotic formula 
\begin{equation}
\label{Minus1}
D \,\,\, \lesssim \,\,\, \big| \,  c \, - \, c_{0} (\alpha) \, \big|^{-1}
\end{equation}

 The scaling law (\ref{Minus1}) differs from the well-known laws in the theory 
of random potentials (or percolation theory). In particular, for a number of 
standard models of random potentials, the following estimate is adopted:
$$D \,\,\, \sim \,\,\, \big| \,  c \, - \, c_{0} (\alpha) \, \big|^{-4/3} $$
(see, e.g., \cite{Stauffer,Essam,Riedel,Trugman}) - though, strictly speaking, 
this law may depend on the specific model.

\vspace{1mm}

 The presence of reflection symmetry imposes significant constraints on the
potentials $\, V \left( {\bf r} , \, \alpha , {\bf a} \right) \, $, even though 
such symmetry is quite common in the physics of two-dimensional systems. 
In general, the absence of this symmetry (while rotational symmetry is maintained), 
probably, brings the properties of the potentials 
$\, V \left( {\bf r} , \, \alpha , {\bf a} \right) \, $
closer to those of random potentials in a plane.

\vspace{1mm}

 In the general case, based on the reasoning in \cite{Superpos}, 
one can derive an estimate
\begin{equation}
\label{Minus5}
D \,\,\, \lesssim \,\,\, \big| \,  c \, - \, c_{0} (\alpha) \, \big|^{-5}
\,\,\, , 
\end{equation}
for the size of closed level lines of potentials 
$\, V \left( {\bf r} , \, \alpha , {\bf a} \right) \, $
that possess rotational symmetry. This estimate, however, appears to be 
a significant overestimation for most potentials encountered in physical 
systems. In particular, it seems to be a gross overestimation for potentials 
with ``good'' analytical properties (characterized by rapidly decaying Fourier 
harmonics), as it would require the Fourier coefficients to decay rather slowly.

\vspace{1mm}

 In this paper, we provide an estimate for the growth in the size of closed 
(bounded) level sets (\ref{VcLevel}) for potentials $\, V_{1} ({\bf r}) \, $
and $\, U ({\bf r}) \, $ that consist of a finite number of Fourier harmonics. 
This estimate takes the form
\begin{equation}
\label{Minus3}
D \,\,\, \lesssim \,\,\, \big| \,  c \, - \, c_{0} (\alpha) \, \big|^{-3} 
\end{equation}
(we will provide more precise statements below), which represents a significant 
refinement of the asymptotic dependence (\ref{Minus5}). At the same time, 
we acknowledge that this estimate may be subject to further refinement - at least 
in a ``statistical'' sense - for the actual potentials encountered in the physics 
of two-dimensional systems. It is also worth noting that the importance of 
investigating the general statistical (geometric and topological) properties 
of quasiperiodic functions with a finite number of Fourier harmonics was pointed 
out by V.I. Arnold (\cite{Arnold1988}).

 In general, approximation by a finite number of Fourier harmonics is a natural 
approach to studying potentials that possess ``good'' analytical properties. 
At the same time, this approximation is exact for many two-dimensional systems 
(in particular, two-dimensional electronic systems and systems of ultracold atoms), 
where artificial ``superpotentials'' are constructed precisely using a finite 
number of harmonics.

\vspace{1mm}

 For the sake of simplicity, we consider here the case where the potential 
$\, V \left( {\bf r} , \, \alpha , {\bf a} \right) \, $ is the sum of the 
potentials $\, V_{1} ({\bf r}) \, $ and $\, V_{2} ({\bf r}) \, $. It is easy 
to see that the results can be readily extended to the case where 
$\, V \left( {\bf r} , \, \alpha , {\bf a} \right) \, $ is a more general linear, 
translationally invariant (i.e., not explicitly dependent on ${\bf r}$) functional 
of $\, V_{1} ({\bf r}) \, $ and $\, V_{2} ({\bf r}) \, $. 

\vspace{1mm}

 Here, we will examine in detail the case of 4th-order symmetry; extending the 
reasoning to the case of 3rd-order (or 6th-order) symmetry is straightforward. 
Furthermore, we will generally assume that we are in a ``generic'' situation. 
This means that certain statements will hold true in all cases, except for specific, 
exceptional scenarios defined by conditions of finite codimension.

\section{Approximation of quasiperiodic potentials by periodic ones and the 
behavior of level lines of $\, V \left( {\bf r} , \, \alpha , {\bf a} \right) \, $ 
near the percolation threshold}
\setcounter{equation}{0}

 We shall assume that the period lattices of $\, V_{1} ({\bf r}) \, $ and 
$\, U ({\bf r}) \, $ are of the form $\, (m, n) \, $ and $\, ( T k , T l ) \, $ 
(where $m, n, k, l \in \mathbb{Z}$), respectively. As is easily seen, 
the periods of $\, V_{2} ({\bf r}) \, $ are then of the form
\begin{equation}
\label{V2Periods}
T \left( k \cos \alpha - l \sin \alpha \, , \,\, 
k \sin \alpha + l \cos \alpha \right)
\end{equation}

 Without loss of generality, we shall assume that $\, T \geq 1 \, $. 
 
 Due to the symmetry of the potentials $\, V_{1} ({\bf r}) \, $ and 
$\, U ({\bf r}) \, $, we can restrict ourselves to rotation angles
$$- \pi / 4 \,\,\, \leq \,\,\, \alpha \,\,\, \leq \,\,\, \pi / 4 $$

 As we have already mentioned, we assume that the potentials 
$\, V_{1} ({\bf r}) \, $ and $\, U ({\bf r}) \, $ contain a finite number of 
Fourier harmonics
$$V_{1} (x, y) \,\,\, = \,\,\, \sum_{|q_{1}|, |q_{2}| \leq Q_{1}}
V_{q_{1}, q_{2}} \, \exp \left( 2 \pi i \left( q_{1} x + q_{2} y \right) \right) 
\,\,\, , $$
$$U (x, y) \,\,\, = \,\,\, \sum_{|q_{1}|, |q_{2}| \leq Q_{2}}
U_{q_{1}, q_{2}} \, \exp {2 \pi i \left( q_{1} x + q_{2} y \right) \over T } $$

 For simplicity, we shall assume here (without loss of generality) that
$\, Q_{1} = Q_{2} = Q \, $. 

\vspace{1mm}

\begin{figure}[t]
\begin{center}
\includegraphics[width=\linewidth]{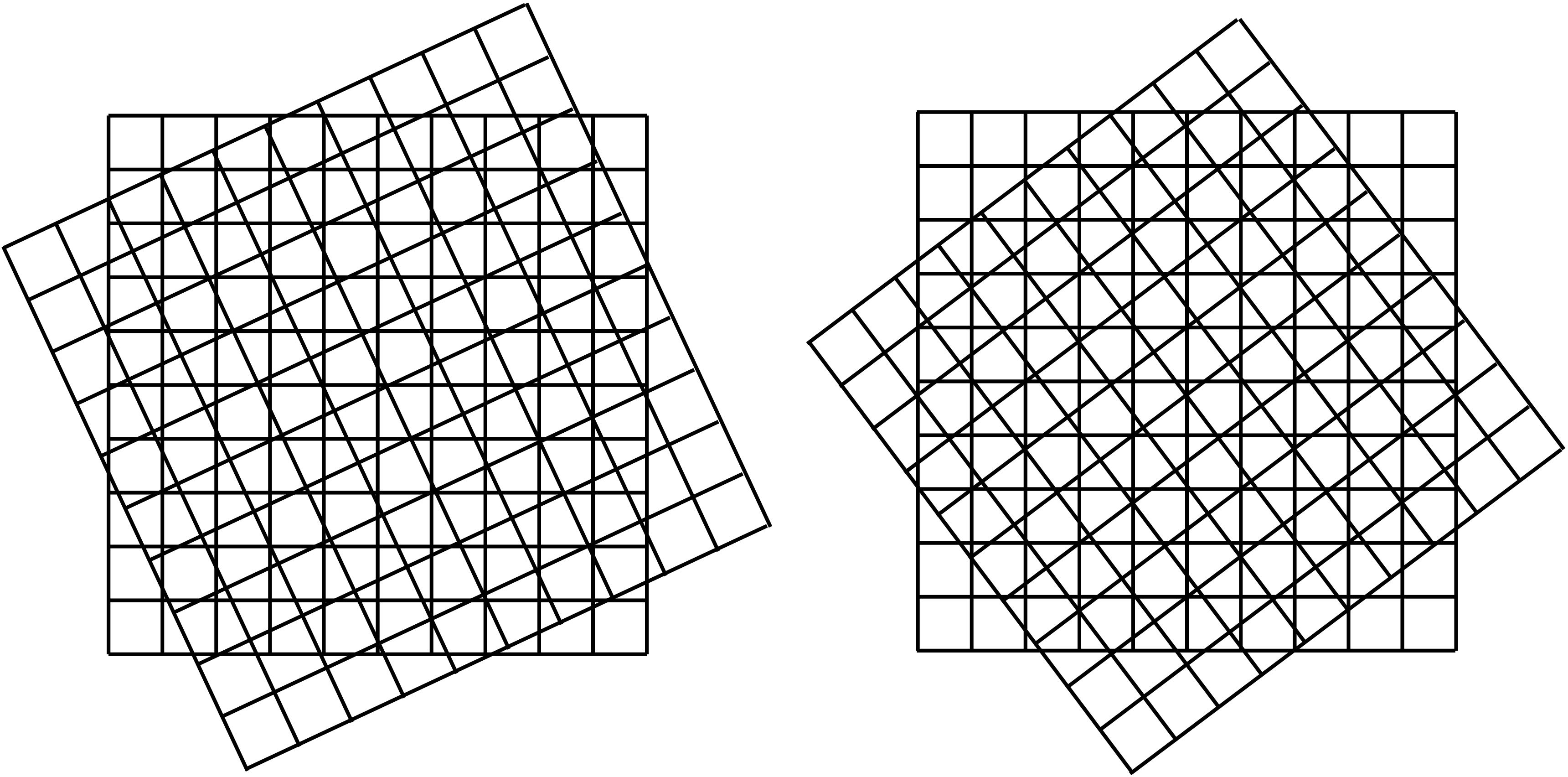}
\end{center}
\caption{Quasiperiodic and periodic potentials
$\, V \left( {\bf r} , \, \alpha , {\bf a} \right) \, $ (schematic representation).}
\label{PeriodicV}
\end{figure}

 As in the works \cite{Superpos,SymSuperPos}, we will employ an approximation of 
quasiperiodic potentials by periodic ones. It is easy to see that periodic potentials 
$\, V \left( {\bf r} , \, \alpha , {\bf a} \right) \, $ arise if the periods of the 
potential $\, U ({\bf r}) \, $ and the rotation angle $\, \alpha \, $ are chosen in 
a specific way (Fig. \ref{PeriodicV}). Naturally, a sufficiently accurate approximation 
of a quasiperiodic potential by a periodic one (with large periods) is possible only 
within a bounded region $\, \Omega \subset \mathbb{R}^{2} \, $.

 In our case, it is convenient to approximate the potential 
$\, V \left( {\bf r} , \, \alpha , {\bf a} \right) \, $ 
within a circle centered at $\, {\bf r}_{0} = (x_{0}, y_{0}) \, $ 
by employing small scalings and rotations of the potential 
$\, V_{2} ({\bf r}) \, $ centered at the point $\, {\bf r}_{0} \, $:
\begin{multline*}
V_{2} ({\bf r}) \,\,\,\,\, \rightarrow \,\,\,\,\, 
{\widehat V}^{2}_{({\bf r}_{0}, \delta \alpha , \delta \lambda )} ({\bf r})
\,\,\,\,\, =  \\
= \,\,\,\,\,  V_{2} \Big( {\bf r}_{0} \, + \, \pi_{- \delta \alpha} \big[ 
(1 - \delta \lambda) ({\bf r} - {\bf r}_{0}) \big] \Big) \,\,\, , 
\end{multline*}
($\delta \alpha , \, \delta \lambda \, \ll \, 1$). 

\vspace{1mm}

\begin{figure}[t]
\begin{center}
\includegraphics[width=0.7\linewidth]{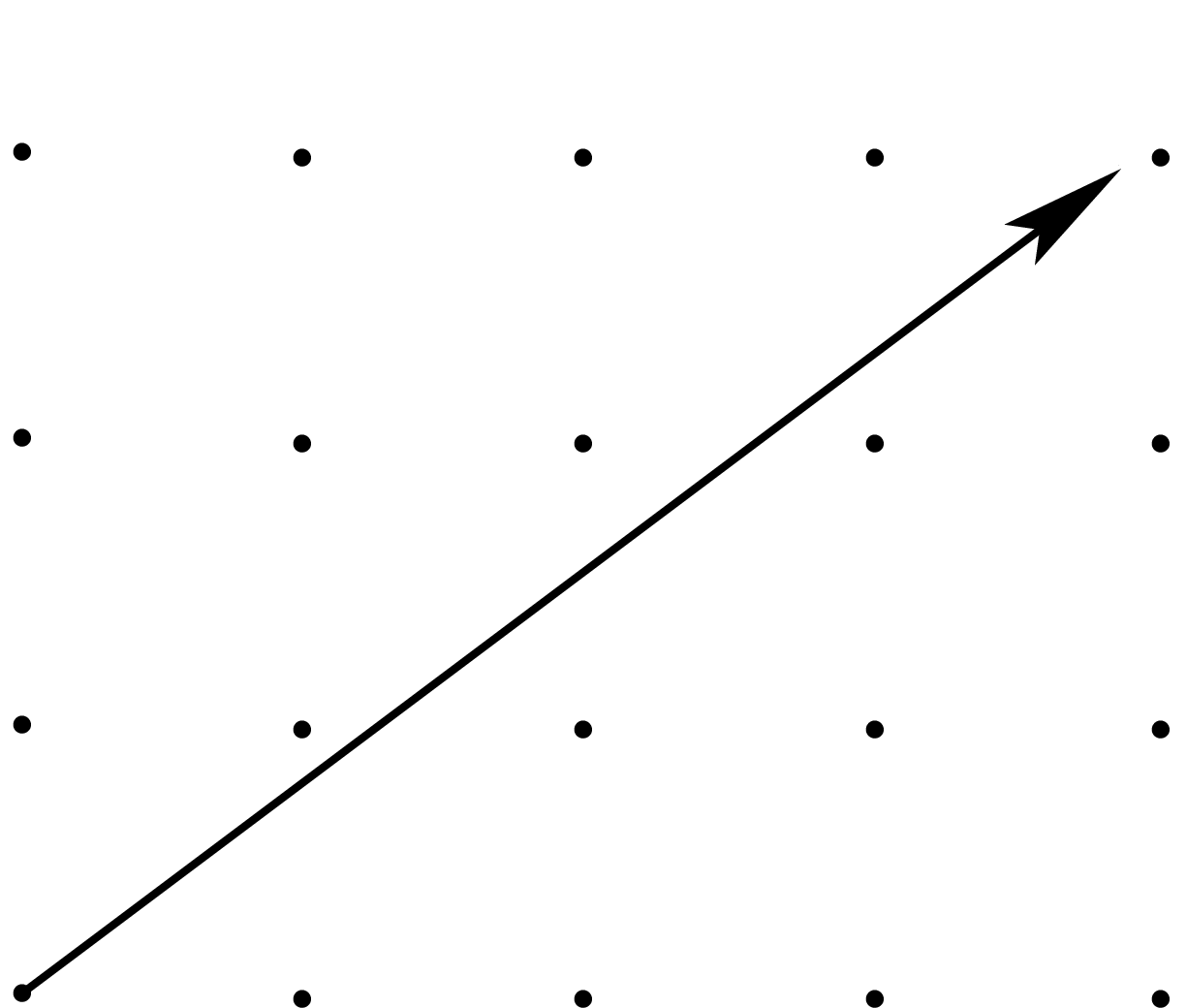}
\end{center}
\caption{The lattice of periods of the potential $\, V_{1} ({\bf r}) \, $ 
and a period of the potential $\, V_{2} ({\bf r}) \, $.}
\label{Approxim}
\end{figure}

 The values $\, \delta \alpha \, $ and $\, \delta \lambda \, $ should naturally 
be chosen such that the potentials $\, V_{1} ({\bf r}) \, $ and 
$\, {\widehat V}^{2}_{({\bf r}_{0}, \delta \alpha , \delta \lambda )} ({\bf r}) \, $
share the same (large) periods. Thus, if this period is the $\, (m, n) \, $
period of the potential $\, V_{1} ({\bf r}) \, $, it must satisfy the conditions
\begin{multline}
\label{mnAppr}
m \,\,\, \simeq \,\,\, T \big( k \cos \alpha \, - \, l \sin \alpha \big) 
\,\,\, , \\
n \,\,\, \simeq \,\,\, T \big( k \sin \alpha \, + \, l \cos \alpha \big)  
\,\,\, ,
\end{multline}
for some integers $\, k \, $ and $\, l \, $ (Fig. \ref{Approxim}).

 To obtain the best approximation for the potential 
$\, V \left( {\bf r} , \, \alpha , {\bf a} \right) \, $, it is natural to choose 
the values $\, (m, n, k, l) \, $ for which the approximations (\ref{mnAppr}) are 
the most accurate. Using standard methods for approximating irrational numbers by 
rational ones, it is straightforward to show (see, e.g., \cite{Superpos}) the 
existence of infinite (increasing) sequences of irreducible quadruplets 
$\, (m_{s}, n_{s}, k_{s}, l_{s}) \, $ such that the quantities 
$\, \delta \alpha_{s} \, $ and $\, \delta \lambda_{s} \, $, corresponding to the 
alignment of the periods (\ref{V2Periods}) and $\, (m, n) \, $, satisfy the condition
$$\delta \alpha_{s} \,\,\, \leq \,\,\, 
{K \over m_{s}^{2} + n_{s}^{2}} \quad , \quad \quad 
\delta \lambda_{s} \,\,\, \leq \,\,\, 
{K \over m_{s}^{2} + n_{s}^{2}} \quad , $$
where $\, K \, $ is some (universal) constant. Without loss of generality, 
we can also assume $\, m_{s} \geq n_{s} \geq 0 \, $. 

 The potential
$${\widehat V}^{(s)}_{{\bf r}_{0}} ({\bf r}) \,\,\, = \,\,\, V_{1} ({\bf r}) 
\,\,\, + \,\,\, 
{\widehat V}^{2}_{({\bf r}_{0}, \delta \alpha_{s} , \delta \lambda_{s} )} ({\bf r}) $$
is periodic with periods $\, (m, n) \, $ and $\, ( - n, m) \, $. Without loss of 
generality, these can be considered the minimal periods of 
$\, {\widehat V}^{(s)}_{{\bf r}_{0}} ({\bf r}) \, $. It is also easy to see that 
the potential $\, {\widehat V}^{(s)}_{{\bf r}_{0}} ({\bf r}) \, $ contains no more 
Fourier harmonics than the potential 
$\, V \left( {\bf r} , \, \alpha , {\bf a} \right) \, $. 

 The potential $\, U ({\bf r}) \, $ is evidently smooth; in particular, its gradient 
satisfies the condition
$$\big| \nabla U ({\bf r}) \big| \,\,\, < \,\,\, {\widehat C} $$
for some constant $\, {\widehat C} \, $. From this, we immediately obtain the relation
$$\big| V \left( {\bf r} , \, \alpha , {\bf a} \right) \, - \, 
{\widehat V}^{(s)}_{{\bf r}_{0}} ({\bf r}) \big| \,\,\, < \,\,\, 
{2 K {\widehat C} R \over m_{s}^{2} + n_{s}^{2}} $$
within a circle of radius $\, R \, $ centered at $\, {\bf r}_{0} \, $. 

 For our purposes, however, we will need to approximate 
$\, V \left( {\bf r} , \, \alpha , {\bf a} \right) \, $ using periodic potentials 
that possess exact rotational symmetry. To obtain such potentials, it suffices to 
shift the potentials 
$\, {\widehat V}^{2}_{({\bf r}_{0}, \delta \alpha_{s} , \delta \lambda_{s} )} ({\bf r}) \, $
by certain vectors $\, \delta {\bf a}_{s} \, $ in the plane $\, \mathbb{R}^{2} \, $
(in order to align two centers of symmetry of the potentials $\, V_{1} ({\bf r}) \, $ and
$\, {\widehat V}^{2}_{({\bf r}_{0}, \delta \alpha_{s} , \delta \lambda_{s} )} ({\bf r})$). 
The minimum value of $\, | \delta {\bf a}_{s} | \, $ required for this decreases as 
$\, s \, $ increases; furthermore, one can employ the estimate (see \cite{Superpos})
$$| \delta {\bf a}_{s} | \,\,\, < \,\,\, {{\rm const} \over m_{s}^{2} + n_{s}^{2}} $$
(note that the common center of symmetry for $\, V_{1} ({\bf r}) \, $ and 
$\, {\widehat V}^{2}_{({\bf r}_{0}, \delta \alpha_{s} , \delta \lambda_{s} )} 
({\bf r} - \delta {\bf a}_{s}) \, $
may be located far from the point $\, {\bf r}_{0}$).

 For the potential
$$V^{(s)}_{{\bf r}_{0}} ({\bf r}) \,\,\, = \,\,\, V_{1} ({\bf r}) \,\,\, + \,\,\, 
{\widehat V}^{2}_{({\bf r}_{0}, \delta \alpha_{s} , \delta \lambda_{s} )} 
({\bf r} - \delta {\bf a}_{s}) $$
we then have
$$\big| {\widehat V}^{(s)}_{{\bf r}_{0}} ({\bf r}) - V^{(s)}_{{\bf r}_{0}} ({\bf r})
\big| \,\,\, < \,\,\, {{\rm const} \cdot {\widehat C} \over m_{s}^{2} + n_{s}^{2}} $$

 Assuming $\, R \, $ to be sufficiently large (in our considerations, 
$\, R \rightarrow \infty$), we can thus formulate the following proposition.

\vspace{2mm}

\noindent
{\bf Proposition 2.1}

\vspace{1mm}

 For any (non-periodic) potential
$\, V \left( {\bf r} , \, \alpha , {\bf a} \right) \, $ and any point
$\, {\bf r}_{0} \in \mathbb{R}^{2} \, $ there exists a sequence of integers
$\, m_{s}, \, n_{s} \, $ ($m_{s} \geq n_{s} \geq 0 \, $,  
$\, m_{s}^{2} + n_{s}^{2} \rightarrow \infty $)  and a sequence of potentials
$\, V^{(s)}_{{\bf r}_{0}} ({\bf r}) \, $ - each periodic with periods
$\, ( m_{s}, n_{s} ) \, $ and $\, ( - n_{s}, m_{s} ) \, $ and possessing exact 
rotational symmetry - such that, within the circle of radius $\, R \, $ 
centered at $\, {\bf r}_{0} \, $, the following condition holds:
$$\big| V \left( {\bf r} , \, \alpha , {\bf a} \right) \, - \, 
V^{(s)}_{{\bf r}_{0}} ({\bf r}) \big| \,\,\, < \,\,\, 
{{\widetilde C} R \over m_{s}^{2} + n_{s}^{2}} $$
(with some constant $\, {\widetilde C} \, $ that is the same for the entire 
family of potentials $\, V \left( {\bf r} , \, \alpha , {\bf a} \right) $). 

\vspace{3mm}

 In what follows, we shall assume throughout that the potentials 
$\, V^{(s)}_{{\bf r}_{0}} ({\bf r}) \, $ are generic. This implies, 
in particular, that their critical points are non-degenerate 
(i.e., they are minima, maxima, or saddle points). Furthermore, we assume 
that non-equivalent critical points of 
$\, V^{(s)}_{{\bf r}_{0}} ({\bf r}) \, $ - that is, those not related by a 
lattice translation - do not lie at the same energy level, unless this is
required by the (rotational) symmetry of the potential
$\, V^{(s)}_{{\bf r}_{0}} ({\bf r}) \, $.

\vspace{1mm}

 Level lines
\begin{equation}
\label{VsLevel}
V^{(s)}_{{\bf r}_{0}} ({\bf r}) \,\,\, = \,\,\, c \,\,\, , 
\end{equation}
that are adjacent to the singular points of the potential 
$\, V^{(s)}_{{\bf r}_{0}} ({\bf r}) \, $ will be referred to as singular level 
lines. The remaining level lines of the potential
$\, V^{(s)}_{{\bf r}_{0}} ({\bf r}) \, $ will be called non-singular or regular.

\vspace{1mm}

\begin{figure}[t]
\begin{center}
\includegraphics[width=0.9\linewidth]{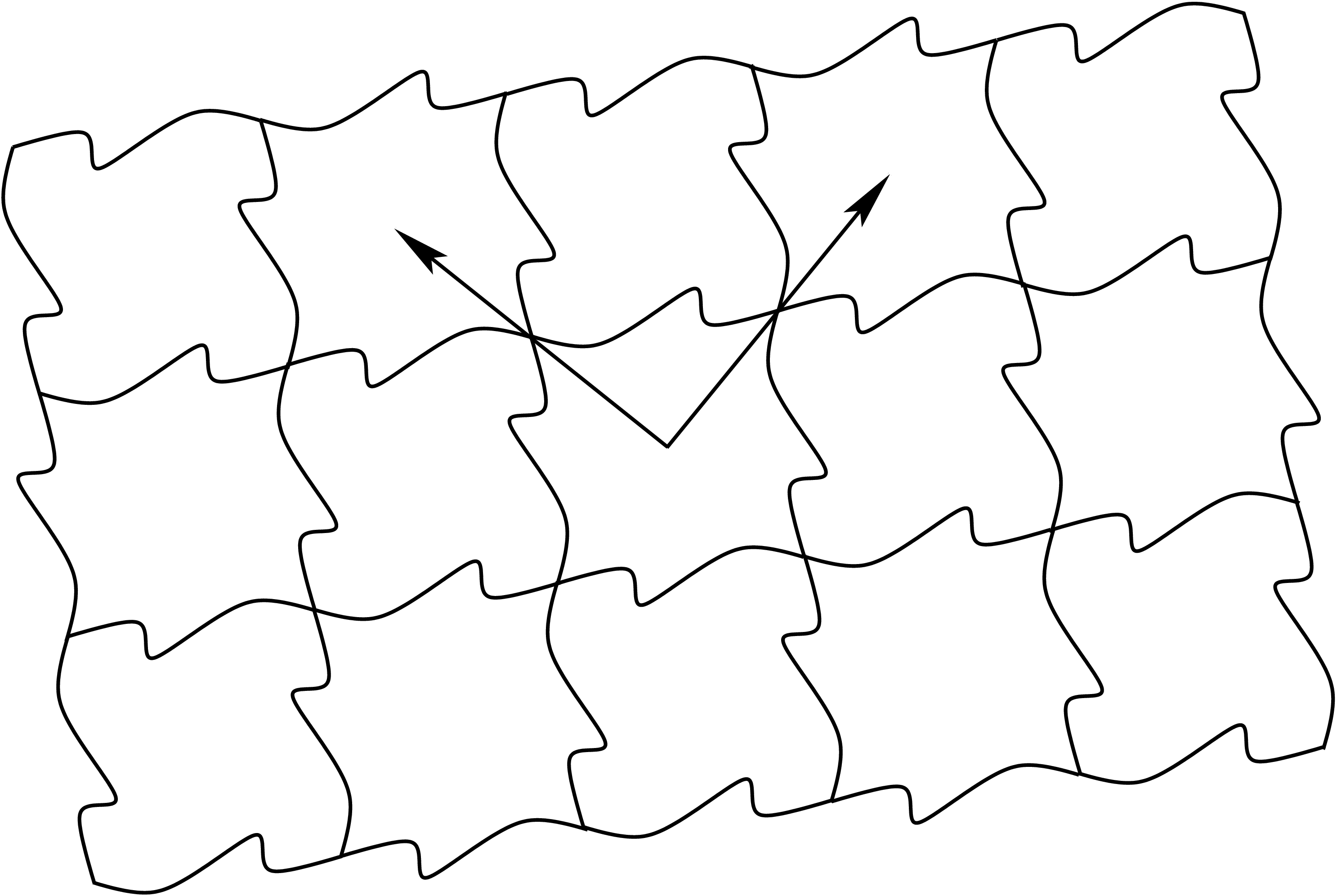}
\end{center}
\caption{A periodic singular net formed by the level lines (\ref{VsLevel}) at the level 
$\, c = c_{s} \, $.} 
\label{SingNet}
\end{figure}

 Unbounded level lines (\ref{VsLevel}) (for periodic potentials possessing exact 
rotational symmetry) exist at a single level, $\, c = c_{s} \, $. For generic 
potentials $\, V^{(s)}_{{\bf r}_{0}} ({\bf r}) \, $, these lines form a symmetric 
net of singular level lines (a singular net) that divides the plane into cells of 
two types - those adjacent to the net on the side of higher values of 
$\, V^{(s)}_{{\bf r}_{0}} ({\bf r}) \, $ and those on the side of lower values 
(Fig. \ref{SingNet}).

 Any two cells of the same type are related by a certain shift by a period of the 
potential $\, V^{(s)}_{{\bf r}_{0}} ({\bf r}) \, $. Each of the cells, evidently, 
possesses 4-fold rotational symmetry.

\vspace{1mm}

 Given a specific topological structure (Fig. \ref{SingNet}), the singular net of 
a potential $\, V^{(s)}_{{\bf r}_{0}} ({\bf r}) \, $ can exhibit a rather complex 
geometry. In particular, the (linear) sizes of its cells may significantly exceed 
the periods of $\, V^{(s)}_{{\bf r}_{0}} ({\bf r}) \, $ (see, e.g., \cite{Superpos}). 
Here, we provide an estimate for the sizes of the cells of the singular net, 
based on our assumptions regarding the potentials $\, V_{1} ({\bf r}) \, $,
$\, V_{2} ({\bf r}) \, $, and $\, V^{(s)}_{{\bf r}_{0}} ({\bf r}) \, $.

\vspace{3mm}

\noindent
{\bf Proposition 2.2}

 For generic potentials $\, V^{(s)}_{{\bf r}_{0}} ({\bf r}) \, $ 
the linear sizes of the cells of the singular net satisfy the condition
\begin{equation}
\label{Dsizes}
D \,\,\, < \,\,\, C^{\prime} \, \left( m_{s}^{2} \, + \, n_{s}^{2} \right)^{3/4}  
\end{equation}
with some constant $\, C^{\prime} \, $ that is the same for the entire 
family of potentials $\, V \left( {\bf r} , \, \alpha , {\bf a} \right) $. 

\vspace{3mm}

 We will provide the proof of Proposition 2.2 in the Appendix.
 
\vspace{3mm}

 As we have already mentioned, the singular net of a potential 
$\, V^{(s)}_{{\bf r}_{0}} ({\bf r}) \, $ arises at a certain isolated level 
$\, c = c_{s} \, $. Furthermore, Propositions 2.1 and 2.2 allow us to make 
another important assertion.

\vspace{3mm}

\noindent
{\bf Proposition 2.3}

\vspace{1mm}

 The sizes of the connected components of (\ref{VcLevel}) do not exceed
$$C^{\prime} \, \left( m_{s}^{2} \, + \, n_{s}^{2} \right)^{3/4} $$
if
$$c \,\,\, \notin \,\,\, \left[ c_{s} - \delta_{s} , \,  c_{s} + \delta_{s}
\right] \,\,\, , $$
where 
$$\delta_{s} \,\,\, = \,\,\,  
{{\widetilde C} C^{\prime}  \over \left( m_{s}^{2} + n_{s}^{2} \right)^{1/4}} $$

\vspace{3mm}

 Indeed, let $\, {\cal D} \, $ be a connected component of the set 
(\ref{VcLevel}) at the level $\, c \, $. Choose a point 
$\, {\bf r}_{0} \in {\cal D} \, $. Suppose that the point $\, {\bf r}_{0} \, $
lies within some cell $\, O \, $ of the singular net of the potential 
$\, V^{(s)}_{{\bf r}_{0}} ({\bf r}) \, $ (assuming the general position case). 
According to Proposition 2.2, the cell $\, O \, $ lies entirely within the 
circle $\, Q_{R} \, $ of radius
$$R \,\,\, = \,\,\, 
C^{\prime} \, \left( m_{s}^{2} \, + \, n_{s}^{2} \right)^{3/4} $$
centered at $\, {\bf r}_{0} \, $. 

 Suppose, for example, that $\, c < c_{s} - \delta_{s} \, $. 
Then, according to Proposition 2.1, the value of the potential 
$\, V^{(s)}_{{\bf r}_{0}} ({\bf r}) \, $ on the part of the component 
$\, {\cal D} \, $ that lies within the circle $\, Q_{R} \, $ does not 
exceed $\, c_{s} \, $; consequently, the component $\, {\cal D} \, $
cannot extend beyond the cell $\, O \, $ (i.e., $\, {\cal D} \, $ does not 
intersect the boundary of $\, O$). 
(The reasoning for $\, c > c_{s} + \delta_{s} \, $ is analogous.)

\vspace{3mm}

 Proposition 2.3, in particular, implies that for any $\, s \, $, 
the following inequality holds:
\begin{equation}
\label{csc0}
\left| c_{0} (\alpha) \, - \, c_{s} \right| \,\,\, < \,\,\, 
\delta_{s} \,\,\, = \,\,\, 
{{\widetilde C} C^{\prime}  \over \left( m_{s}^{2} + n_{s}^{2} \right)^{1/4}} 
\end{equation}

\vspace{2mm}

 From Proposition 2.1 and the relations (\ref{csc0}), we, in turn, 
obtain the following proposition.

\vspace{3mm}

 For any (non-periodic) potential 
$\, V \left( {\bf r} , \, \alpha , {\bf a} \right) \, $, 
there exists an increasing sequence of integers $\, (m_{s}, n_{s}) \, $
such that the sizes of the connected components of the level set 
(\ref{VcLevel}) do not exceed
$$C^{\prime} \, \left( m_{s}^{2} \, + \, n_{s}^{2} \right)^{3/4} $$
if
$$\big| c_{0} (\alpha) \, - \, c \big| \,\,\, > \,\,\, 2 \delta_{s} 
\,\,\, = \,\,\, 
{2 {\widetilde C} C^{\prime}  \over \left( m_{s}^{2} + n_{s}^{2} \right)^{1/4}} $$

\vspace{3mm}

 The statement above can also be formulated in the following way.

\vspace{3mm}

\noindent
{\bf Proposition 2.4}

\vspace{1mm}

 For any (non-periodic) potential 
$\, V \left( {\bf r} , \, \alpha , {\bf a} \right) \, $,
there exists a decreasing sequence $\, \Delta_{s} \rightarrow 0 \, $
such that the sizes of the connected components of the level set 
(\ref{VcLevel}) do not exceed $\, C \Delta_{s}^{-3} \, $ if
$$\big| c_{0} (\alpha) \, - \, c \big| \,\,\, > \,\,\, \Delta_{s} $$
(with some constant $\, C \, $ that is the same for the entire family 
of potentials $\, V \left( {\bf r} , \, \alpha , {\bf a} \right) $).

\vspace{3mm}

 Proposition 2.4 provides a more precise formulation of the relation 
(\ref{Minus3}). This formulation can be somewhat ``regularized'', 
based on a specific regularity in the growth of the numbers 
$\, (m_{s}, n_{s}) \, $ for ``almost all'' non-periodic potentials 
$\, V \left( {\bf r} , \, \alpha , {\bf a} \right) \, $
(see, e.g., \cite{SymSuperPos}). Here, however, we will not examine 
the corresponding refinements in detail.

\section{Appendix}
\setcounter{equation}{0}

 Here we provide a proof of Proposition 2.2, based on the structure of 
the potentials $\, V^{(s)}_{{\bf r}_{0}} ({\bf r}) \, $. 

\vspace{2mm}

 For each $\, s \, $, let
$$ {\bf t}^{(s)}_{1} \,\, = \,\, (m_{s}, \, n_{s}) \,\,\, , \quad
{\bf t}^{(s)}_{2} \,\, = \,\, (- n_{s}, \, m_{s}) $$
be the minimal periods of the potential 
$\, V^{(s)}_{{\bf r}_{0}} ({\bf r}) \, $, and
$$L^{(s)} \,\,\, = \,\,\, \left\{ p \, {\bf t}^{(s)}_{1} \,\, + \,\, 
q \, {\bf t}^{(s)}_{2} \,\,\, , \quad p, q \in \mathbb{Z} \right\} $$
be the integer lattice corresponding to the potential
$\, V^{(s)}_{{\bf r}_{0}} ({\bf r}) \, $.

\vspace{2mm}

 We will need here the following simple lemma.
 
\vspace{2mm}

\noindent
{\bf Lemma 3.1}

 Let $\, {\bf g} \subset \mathbb{R}^{2} \, $ be a straight line segment 
in $\, \mathbb{R}^{2} \, $ with endpoints at $\, (x, y) \, $ 
and $\, (x + m_{s} , \,\, y + n_{s} )$. 
Then (for sufficiently large $\, s$), the restriction 
$$V^{(s)}_{{\bf r}_{0}} ({\bf r}) \big|_{\bf g} $$
takes on any of its values at most $\, 4 Q \sqrt{ m_{s}^{2} + n_{s}^{2}} \, $ times.  
(The same holds for the orthogonal line segment with endpoints at 
$\, (x, y) \, $ and $\, (x - n_{s} , \,\, y + m_{s})$).

\vspace{1mm}

\noindent
{\bf Remark.}

 Here, we assume that the generic potential 
$\, V^{(s)}_{{\bf r}_{0}} ({\bf r}) \, $
is not identically constant on the segment $\, {\bf g} \, $. 

\vspace{1mm}

\noindent
Proof.

 The restriction of $\, V^{(s)}_{{\bf r}_{0}} ({\bf r}) \, $
to any straight line $\, b \subset \mathbb{R}^{2} \, $ parallel to 
$\, (m_{s} , n_{s} ) \, $ is a periodic function (with period 
$\sqrt{m_{s}^{2} + n_{s}^{2}}$) that contains a finite number of 
Fourier harmonics.

 The largest wavenumber in its Fourier expansion is evidently 
determined by the maximum of the quantities
$$Q \,\,\, { | m_{s} | + | n_{s} | \over \sqrt{ m_{s}^{2} + n_{s}^{2}}} 
\quad \quad  \text{and}  \quad \quad 
{Q \over T} \,\, { \big( | k_{s} | + | l_{s} | \big) \, 
\big( 1 - \delta \lambda_{s} \big) \over \sqrt{ k_{s}^{2} + l_{s}^{2}}} 
\,\,\, , $$
that do not exceed
$$Q \, {\sqrt{2 (m_{s}^{2} + n_{s}^{2})} \over \sqrt{ m_{s}^{2} + n_{s}^{2}}}
\,\,\, = \,\,\, \sqrt{2} \, Q  \,\,\, < \,\,\, 2 Q $$
and
$${Q \big( 1 - \delta \lambda_{s} \big) \over T} \, 
{\sqrt{2 (k_{s}^{2} + l_{s}^{2})} \over \sqrt{ k_{s}^{2} + l_{s}^{2}}}
\,\,\, = \,\,\, {\sqrt{2} Q \big( 1 - \delta \lambda_{s} \big) \over T} 
\,\,\, < \,\,\, 2 Q $$
(for sufficiently small $\delta \lambda_{s}$), respectively.

 Given that the length of the segment $\, {\bf g} \, $ is equal to 
$\, \sqrt{ m_{s}^{2} + n_{s}^{2}} \, $, it can be seen that the statement of 
the lemma is a well-known result from the theory of trigonometric polynomials.

\hfill{Lemma 3.1 is proved.}

\vspace{3mm}

\begin{figure}[t]
\begin{center}
\includegraphics[width=0.9\linewidth]{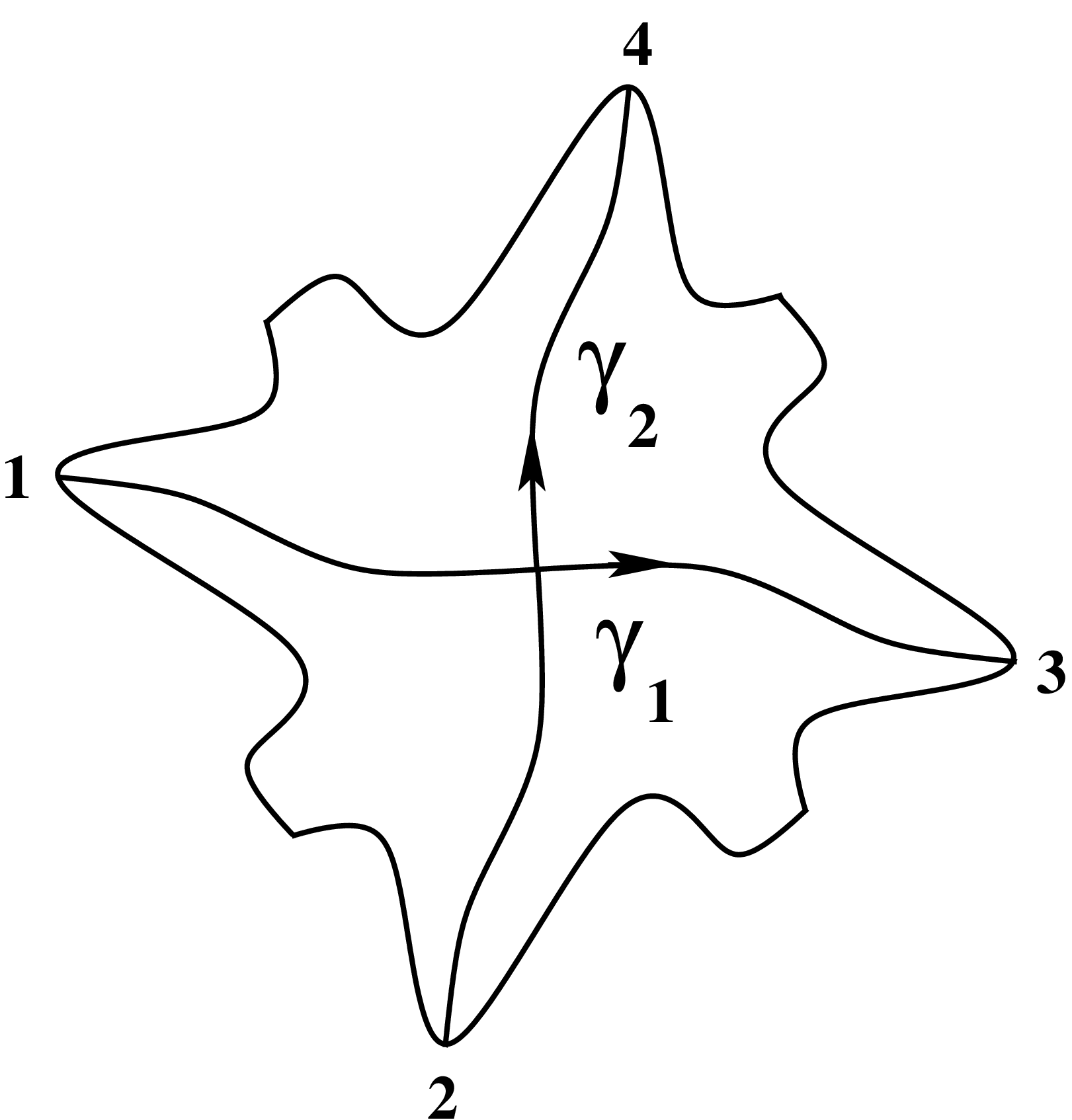}
\end{center}
\caption{Points on the boundary of a cell of the singular net that 
are furthest from its center, and paths $\, \gamma_{1} \, $, 
$\, \gamma_{2} \, $ connecting pairs of opposite points. }
\label{Congl}
\end{figure}

 In a cell of the singular net, consider four points on its boundary that are 
furthest from its center and map onto one another under $90^{\circ}$ rotations 
(Fig. \ref{Congl}). Connect these points to the center using four non-intersecting 
paths, which also map onto one another under $90^{\circ}$ rotations (it is easy 
to show that such paths exist). Consider the oriented paths $\, \gamma_{1} \, $ 
and $\, \gamma_{2} \, $ that connect points 1 and 3, and points 2 and 4, 
respectively (Fig. \ref{Congl}). The paths $\, \gamma_{1} \, $ and 
$\, \gamma_{2} \, $ map onto each other under a $90^{\circ}$ rotation and 
have exactly one point of intersection. Clearly, the distance between points 1 
and 3 (as well as between points 2 and 4) is equal to the diameter $\, D \, $ of 
the cell.

\vspace{1mm}

 The conglomerate $\, \gamma_{1} \cup \gamma_{2} \, $ does not intersect 
any of its integer shifts
$$\gamma_{1} \cup \gamma_{2} \,\, + \,\, p \, {\bf t}^{(s)}_{1} \,\, + \,\, 
q \, {\bf t}^{(s)}_{2} \,\,\, , \quad p, q \in \mathbb{Z} $$

 This, in particular, means that $\, \gamma_{1} \cup \gamma_{2} \, $ is 
mapped bijectively onto its image in the torus $\, \mathbb{T}^{2} \, $ under 
the covering map
\begin{equation}
\label{Covering}
P : \quad \mathbb{R}^{2} \,\,\, \rightarrow \,\,\, \mathbb{T}^{2} \,\, = \,\, 
\mathbb{R}^{2} \big/ L^{(s)} 
\end{equation}

 As is not difficult to show, among the integer shifts $1^{\prime}$ of the 
point $1$ we can choose one such that

\vspace{2mm}

\begin{figure}[t]
\begin{center}
\includegraphics[width=\linewidth]{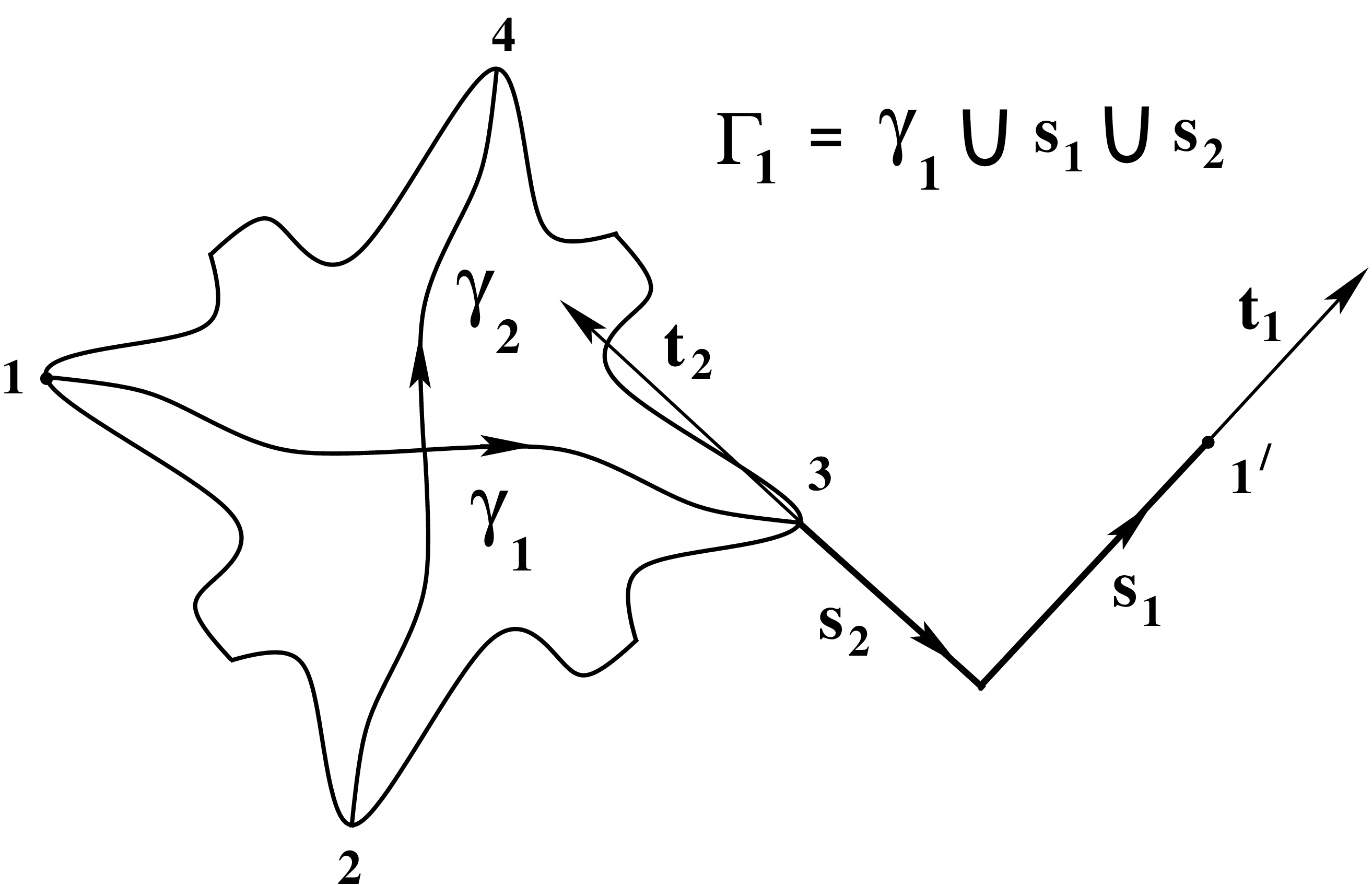}
\end{center}
\caption{The path $\, \Gamma_{1} \, $, connecting points $1$ and $1^{\prime}$ 
that differ by a shift equal to a (certain) period of the potential 
$\, V^{(s)}_{{\bf r}_{0}} ({\bf r}) \, $.}
\label{PutGamma1}
\end{figure}

\noindent
1) Points $3$ and $1^{\prime}$ can be connected by straight-line segments 
$\, {\bf s}_{1} \, $ and $\, {\bf s}_{2} \, $, which are parts of the periods 
$\, {\bf t}^{(s)}_{1} \, $ and $\, {\bf t}^{(s)}_{2} \, $ (Fig. \ref{PutGamma1});

\vspace{2mm}

\noindent
2) The distance between the endpoints of the path
$$\Gamma_{1} \,\,\, = \,\,\, \gamma_{1} \cup {\bf s}_{1} \cup {\bf s}_{2} 
\,\,\, = \,\,\, \gamma_{1} \cup {\bf s} \quad \quad \left( 
{\bf s} \, = \, {\bf s}_{1} \cup {\bf s}_{2} \right) $$
is no less than the distance between points 1 and 3 (i.e., the diameter $D$).

\vspace{2mm}

 Let us choose the orientation of $\, \Gamma_{1} \, $ that coincides with 
the orientation of $\, \gamma_{1} \, $. The path $\, \Gamma_{1} \, $
connects points in $\, \mathbb{R}^{2} \, $ that differ by an integer shift, 
and it maps to a closed loop on the torus  $\, \mathbb{T}^{2} \, $ under 
the covering map (\ref{Covering}).

\vspace{1mm}

 By performing an analogous operation with the path $\, \gamma_{2} \, $, 
we obtain the path
$$\Gamma_{2} \,\,\, = \,\,\, \gamma_{2} \cup {\bf s}^{\prime} \,\,\, , $$
which is derived from $\, \Gamma_{1} \, $ by a $90^{\circ}$ rotation.

\vspace{1mm}

 Let the points $1$ and $1^{\prime}$ differ by the vector
$$\, p \, {\bf t}^{(s)}_{1} \,\, + \,\, q \, {\bf t}^{(s)}_{2} \,\,\, , 
\quad p, q \in \mathbb{Z} $$

 As is easy to see, the distance between $1$ and $1^{\prime}$ is then equal to
$$\sqrt{ p^{2} + q^{2}} \,\, \sqrt{m_{s}^{2} + n_{s}^{2}} $$

 Under the mapping (\ref{Covering}), the image $\, P \left( \Gamma_{1} \right) \, $ 
is the cycle
$$p \, {\bf a}^{(s)}_{1} \,\, + \,\, q \, {\bf a}^{(s)}_{2} \,\,\, , $$
where $\, {\bf a}^{(s)}_{1} \, $ and $\, {\bf a}^{(s)}_{2} \, $ are the standard 
basis of the one-dimensional homology of the torus $\, \mathbb{T}^{2} \, $. 

\vspace{1mm}

 Similarly, $\, P \left( \Gamma_{2} \right) \, $ represents the cycle
$$- \, q \, {\bf a}^{(s)}_{1} \,\, + \,\, p \, {\bf a}^{(s)}_{2} \,\,\, , $$
and the intersection index 
\begin{equation}
\label{Index}
P \left( \Gamma_{1} \right) \, \circ \, P \left( \Gamma_{2} \right) 
\end{equation}
is equal to $\, p^{2} + q^{2} \, $. 

\vspace{1mm}

 At the same time, the intersection index (\ref{Index}) can also be calculated 
directly in the plane $\, \mathbb{R}^{2} \, $. To do this, we define an operation 
$\, {\hat \circ} \, $ as the signed number of intersections of two arbitrary 
oriented curves in the plane. The intersection index (\ref{Index}) can then be 
defined by the formula
\begin{multline*}
P \left( \Gamma_{1} \right) \, \circ \, P \left( \Gamma_{2} \right) 
\,\,\, =  \\
= \,\,\, \Gamma_{1} \, {\hat \circ} \, 
\bigcup_{k,l \in \mathbb{Z}} \left( \Gamma_{2} \, + \, 
k \, {\bf t}^{(s)}_{1} \,\, + \,\, l \, {\bf t}^{(s)}_{2} \right) 
\,\,\, =  \\
= \,\,\, \sum_{k,l \in \mathbb{Z}} \,
\Gamma_{1} \, {\hat \circ} \, \left( \Gamma_{2} \, + \, 
k \, {\bf t}^{(s)}_{1} \,\, + \,\, l \, {\bf t}^{(s)}_{2} \right) 
\end{multline*}
(sum over all integer shifts of the path $\Gamma_{2}$).

\vspace{1mm}

 The latter sum, in turn, can be expressed as
\begin{multline}
\label{BigSum}
\sum_{k,l \in \mathbb{Z}} \,
\gamma_{1} \, {\hat \circ} \, \left( \gamma_{2} \, + \, 
k \, {\bf t}^{(s)}_{1} \,\, + \,\, l \, {\bf t}^{(s)}_{2} \right) 
\,\,\, + \\
+ \,\,\, \sum_{k,l \in \mathbb{Z}} \,
\gamma_{1} \, {\hat \circ} \, \left( {\bf s}^{\prime} \, + \, 
k \, {\bf t}^{(s)}_{1} \,\, + \,\, l \, {\bf t}^{(s)}_{2} \right) 
\,\,\, + \\
+ \,\,\, \sum_{k,l \in \mathbb{Z}} \,
{\bf s} \, {\hat \circ} \, \left( \gamma_{2} \, + \, 
k \, {\bf t}^{(s)}_{1} \,\, + \,\, l \, {\bf t}^{(s)}_{2} \right) 
\,\,\, + \\
+ \,\,\, \sum_{k,l \in \mathbb{Z}} \,
{\bf s} \, {\hat \circ} \, \left( {\bf s}^{\prime} \, + \, 
k \, {\bf t}^{(s)}_{1} \,\, + \,\, l \, {\bf t}^{(s)}_{2} \right) 
\end{multline}

\vspace{1mm}

 By construction, the path $\, \gamma_{1} \, $ intersects only one (the zero) 
integer shift of $\, \gamma_{2} \, $, and the first term in the sum 
(\ref{BigSum}) is equal to 1.

\vspace{1mm}

 It is not difficult to show that the second and third terms of the sum 
(\ref{BigSum}) are equal to each other. Let us examine the third term of 
the sum (\ref{BigSum}) in more detail.

\vspace{1mm}

\begin{figure}[t]
\begin{center}
\includegraphics[width=\linewidth]{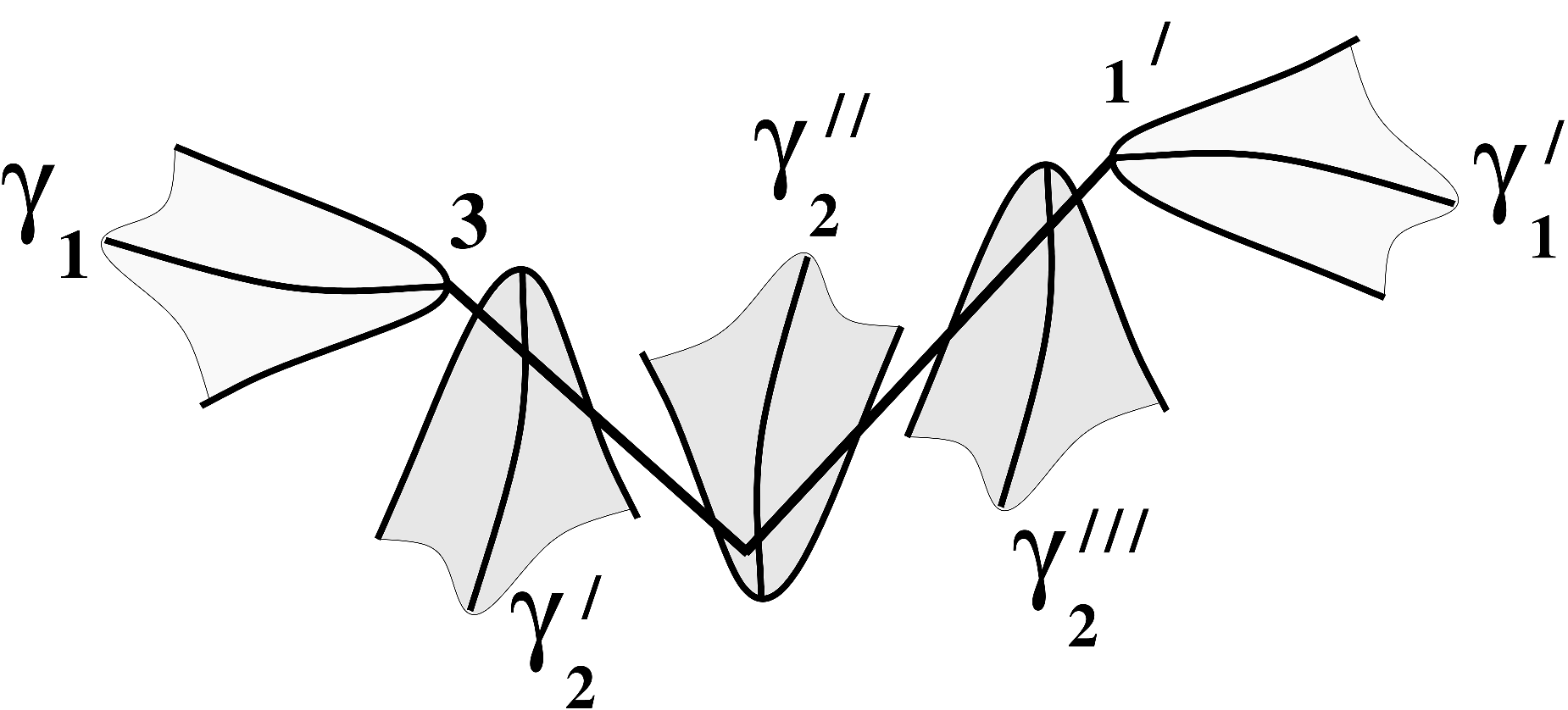}
\end{center}
\caption{The intersections of the polygonal line $\, {\bf s} \, $ with 
integer shifts of the curve $\, \gamma_{2} \, $.}
\label{sgammaint}
\end{figure}

 By examining the intersections of $\, {\bf s} \, $ (which is adjacent to the 
paths $\gamma_{1}$ and $\gamma_{1}^{\prime}$) with integer shifts of 
$\, \gamma_{2} \, $, one can observe that none of the shifts of $\, \gamma_{2} \, $ 
can ``go around'' $\, \gamma_{1} \, $ or $\, \gamma_{1}^{\prime} \, $ 
(given their maximal sizes). Consequently, the total intersection index 
of $\, {\bf s} \, $ with any of the shifts $\, \gamma_{2} \, $ can only be 
$-1$, $0$, or $1$ (Fig. \ref{sgammaint}). The value of the third term in the 
sum (\ref{BigSum}) is, therefore, bounded by the number of intersections 
of $\, {\bf s} \, $ with {\it different} integer shifts of the path 
$\, \gamma_{2} \, $.

 Each intersection of the polygonal line $\, {\bf s} \, $ with an integer 
shift of $\, \gamma_{2} \, $ means its intersection with the corresponding 
cell of the singular net of the potential $\, V^{(s)}_{{\bf r}_{0}} ({\bf r}) \, $. 
This, in turn, implies that the value 
$\, V^{(s)}_{{\bf r}_{0}} ({\bf r}) = c_{s} \, $ is attained at at least 
two points where the line $\, {\bf s} \, $ crosses the boundary of this cell. 
According to Lemma 3.1, the total number of such points along the path 
$\, {\bf s} \, $ cannot exceed $\, 8 Q \sqrt{ m_{s}^{2} + n_{s}^{2}} \, $. 
Consequently, the third term in (\ref{BigSum}) (as well as the second term) 
is bounded by the value $\, 4 Q \sqrt{ m_{s}^{2} + n_{s}^{2}} \, $.

\vspace{1mm}

\begin{figure}[t]
\begin{center}
\includegraphics[width=0.7\linewidth]{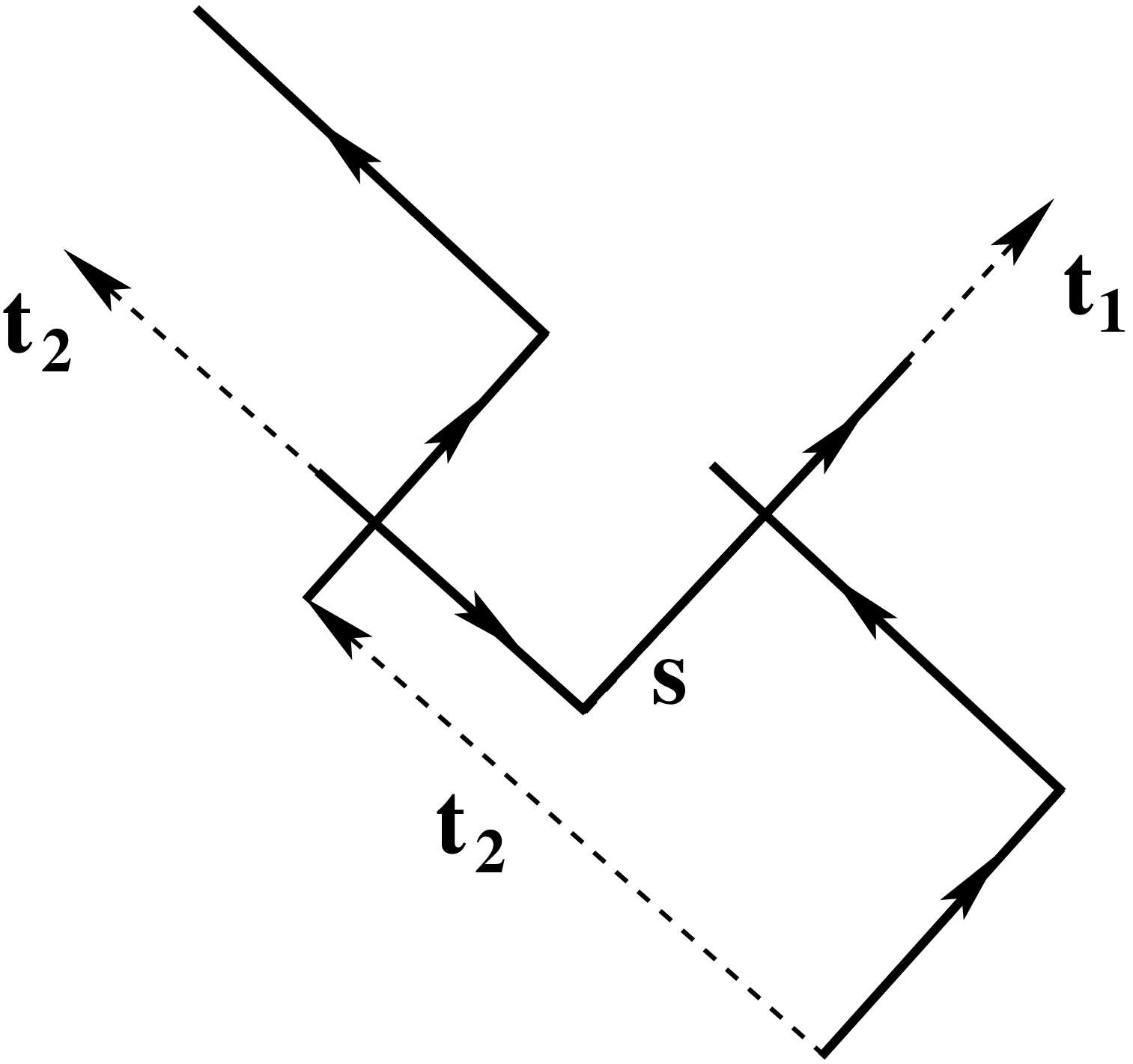}
\end{center}
\caption{A polygonal line $\, {\bf s} \, $ that intersects two integer 
shifts of the polygonal line  $\, {\bf s}^{\prime} \, $.}
\label{sprimesint}
\end{figure}

 As for the last term in (\ref{BigSum}), it is easy to see that the 
intersection of $\, {\bf s} \, $ with an integer shift of 
$\, {\bf s}^{\prime} \, $ (in general position) can occur at only one point; 
furthermore, the number of such shifts of $\, {\bf s}^{\prime} \, $ that 
intersect $\, {\bf s} \, $ cannot exceed 2 (Fig. \ref{sprimesint}).

\vspace{1mm}

 We thus obtain
$$ p^{2} + q^{2} \,\,\, \leq \,\,\, 
8 Q \sqrt{ m_{s}^{2} + n_{s}^{2}} \, + \, 3 $$
and
\begin{multline*}
D^{2} \,\,\, \leq \,\,\, \left( p^{2} + q^{2} \right) 
\left( m_{s}^{2} + n_{s}^{2} \right) \,\,\, \leq  \\
\leq  \,\,\, \left( 8 Q \sqrt{ m_{s}^{2} + n_{s}^{2}} \, + \, 3 \right)
\left( m_{s}^{2} + n_{s}^{2} \right) \,\,\, , 
\end{multline*}
which, evidently, entails (\ref{Dsizes}).

\section{Conclusion}
\setcounter{equation}{0}

 This paper analyzes the behavior of the level lines of bilayer potentials 
that possess rotational quasicrystalline symmetry. As an approximation for 
potentials with ``good'' analytical properties, the study employs potentials 
consisting of a finite number of Fourier harmonics. The research yields 
constraints on the growth of the sizes of the level lines of such potentials
near the percolation threshold, which allows for a comparison of their 
behavior with the percolation properties of potentials belonging to other 
classes.


\begin{thebibliography}{99}

\bibitem{LePiunSad} 
Le Thang Tu Quoc, S. A. Piunikhin, V. A. Sadov,
The geometry of quasicrystals, 
{\it Russian Mathematical Surveys} {\bf 48} : 1 (1993),
37–100

\vspace{1mm}

\bibitem{MultValAnMorseTheory} 
S.P. Novikov,
The Hamiltonian formalism and a many-valued analogue of  
Morse theory, {\it Russian Math. Surveys} {\bf 37} (5), 
1-56 (1982).

\vspace{1mm}

\bibitem{zorich1} 
A.V. Zorich,
A problem of Novikov on the semiclassical motion 
of an electron in a uniform almost rational magnetic field.,  
{\it Russian Math. Surveys} {\bf 39} (5), 287-288 (1984).

\vspace{1mm}

\bibitem{dynn1992} 
I.A. Dynnikov,
Proof of S.P. Novikov's conjecture for the case of 
small perturbations of rational magnetic fields,  
{\it Russian Math. Surveys} {\bf 47}:3, 172-173 (1992).

\vspace{1mm}

\bibitem{Tsarev} 
S.P. Tsarev, Seminar talks, 1992-1993

\vspace{1mm}

\bibitem{dynn1} 
I.A. Dynnikov,
Proof of S.P. Novikov's conjecture on
the semiclassical motion of an electron,  
{\it Math. Notes} {\bf 53}:5, 495-501 (1993).

\vspace{1mm}

\bibitem{zorich2} A.V. Zorich.,
Proc. ``Geometric Study of Foliations''.,
(Tokyo, November 1993) / ed. T.Mizutani et al.
Singapore: World Scientific, 479-498 (1994).

\vspace{1mm}

\bibitem{DynnBuDA} I.A. Dynnikov.,
Surfaces in 3-torus: geometry of plane sections., 
Proc. of ECM2, BuDA, 1996.

\vspace{1mm}

\bibitem{dynn2} I.A. Dynnikov, 
Amer. Math. Soc. Transl. Ser. 2,
Vol. 179, AMS, Providence, RI (1997), p. 45.

\vspace{1mm}

\bibitem{dynn3} 
I.A. Dynnikov,
The geometry of stability regions in Novikov's problem on the
semiclassical motion of an electron,  
{\it Russian Math. Surveys} {\bf 54}:1, 21-59 (1999).

\vspace{1mm}

\bibitem{PismaZhETF}
S.P. Novikov, A.Y. Maltsev,
Topological quantum characteristics observed in the 
investigation of the conductivity in normal metals,  
{\it JETP Letters} {\bf 63} (10), 855-860 (1996).

\vspace{1mm}

\bibitem{ZhETF1997}
A.Ya. Maltsev, 
Anomalous behavior of the electrical conductivity tensor in strong 
magnetic fields, J. Exp. Theor. Phys. {\bf 85} (5), 934-942 (1997)

\vspace{1mm}

\bibitem{UFN} 
S.P. Novikov, A.Y. Maltsev,
Topological phenomena in normal metals,  
{\it Physics-Uspekhi} {\bf 41}:3, 231-239 (1998).

\vspace{1mm}

\bibitem{BullBrazMathSoc}
A.Ya. Maltsev and S.P. Novikov,
Quasiperiodic functions and dynamical systems in quantum solid state physics, 
Bulletin of Braz. Math. Society, New Series {\bf 34}, 171 (2003).

\vspace{1mm}

\bibitem{StatPhys}
A.Ya. Maltsev and S.P. Novikov, 
Dynamical Systems, Topology, and Conductivity in Normal Metals,
J. Stat. Phys. {\bf 115}, 31 (2004).

\vspace{1mm}

\bibitem{DeLeoPhysLettA} R. De Leo,
Topological effects in the magnetoresistance of Au and Ag,
Phys. Lett. A {\bf 332}, 469 (2004)

\vspace{1mm}

\bibitem{DeLeoPhysB} R. De Leo,
First-principles generation of stereographic maps for high-field 
magnetoresistance in normal metals: An application to Au and Ag, 
Physica B: Condensed Matter {\bf 362} (1–4), 62 (2005)

\vspace{1mm}

\bibitem{TrMIAN} 
A.Ya. Maltsev, S.P. Novikov,
The Theory of Closed 1-Forms, Levels of Quasiperiodic Functions 
and Transport Phenomena in Electron Systems., 
{\it Proceedings of the Steklov Institute of Mathematics}
{\bf 302}, 279-297 (2018).

\vspace{1mm}

\bibitem{DeLeo2017}
Roberto De Leo,
A survey on quasiperiodic topology,
Advanced Mathematical Methods in Biosciences \& Applications,
Springer, Eds. F. Berezovskaya and B. Toni, (2018)., 
arXiv:1711.01716

\vspace{1mm}

\bibitem{zorich3}
Anton Zorich.,
How do the leaves of closed 1-form wind around a surface.,
``Pseudoperiodic Topology'',
V.I.Arnold, M.Kontsevich, A.Zorich (eds.),
Translations of the AMS, Ser. 2, vol. 197, AMS,
Providence, RI, 1999, 135-178.

\vspace{1mm}

\bibitem{DeLeo1} 
R. De Leo,
Existence and measure of ergodic leaves in Novikov's problem
on the semiclassical motion of an electron.,  
{\it Russian Math. Surveys} {\bf 55}:1 (2000), 166-168.

\vspace{1mm}

\bibitem{DeLeo2} 
R. De Leo,
Characterization of the set of ``ergodic directions'' in Novikov's
problem of quasi-electron orbits in normal metals.,  
{\it Russian Math. Surveys} {\bf 58}:5 (2003), 1042-1043.

\vspace{1mm}

\bibitem{DeLeo3} R. De Leo.,
Topology of plane sections of periodic polyhedra with an application 
to the Truncated Octahedron., 
{\it Experimental Mathematics} {\bf 15}:1 (2006), 109-124.

\vspace{1mm}

\bibitem{DeLeoDynnikov1} 
R. De Leo, I.A. Dynnikov,   
An example of a fractal set of plane directions having chaotic
intersections with a fixed 3-periodic surface.,  
{\it Russian Math. Surveys} {\bf 62}:5 (2007), 990-992.

\vspace{1mm}

\bibitem{Dynnikov2008} I.A. Dynnikov,
Interval identification systems and plane sections of 3-periodic surfaces,
{\it Proceedings of the Steklov Institute of Mathematics}
{\bf 263}, 65-77 (2008).

\vspace{1mm}

\bibitem{DeLeoDynnikov2} R. De Leo, I.A. Dynnikov.,
Geometry of plane sections of the infinite regular skew polyhedron
$\{ 4, \, 6 \, | \, 4 \}$., 
{\it Geom. Dedicata} {\bf 138}:1 (2009), 51-67.

\vspace{1mm}

\bibitem{Skripchenko1} A. Skripchenko.,
Symmetric interval identification systems of order three.,
{\it Discrete Contin. Dyn. Sys.} {\bf 32}:2 (2012), 643-656.

\vspace{1mm}

\bibitem{Skripchenko2} A. Skripchenko.,
On connectedness of chaotic sections of some 3-periodic surfaces.,
{\it Ann. Glob. Anal. Geom.} {\bf 43} (2013), 253-271.

\vspace{1mm}

\bibitem{DynnSkrip1} I. Dynnikov, A. Skripchenko.,
On typical leaves of a measured foliated 2-complex of thin type., 
Topology, Geometry, Integrable Systems, and Mathematical Physics: 
Novikov's Seminar 2012-2014, Advances in the Mathematical Sciences., 
Amer. Math. Soc. Transl. Ser. 2, 234, eds. V.M. Buchstaber, 
B.A. Dubrovin, I.M. Krichever, Amer. Math. Soc., Providence,
RI, 2014, 173-200, arXiv: 1309.4884

\vspace{1mm}

\bibitem{DynnSkrip2} I. Dynnikov, A. Skripchenko.,
Symmetric band complexes of thin type and chaotic sections which 
are not actually chaotic., {\it Trans. Moscow Math. Soc.}, Vol. 76, 
no. 2, 2015, 287-308.

\vspace{1mm}

\bibitem{AvilaHubSkrip1}
A. Avila, P. Hubert, A. Skripchenko.,
Diffusion for chaotic plane sections of 3-periodic surfaces.,
{\it Inventiones mathematicae}, October 2016, Volume 206, 
Issue 1, pp 109–146. 

\vspace{1mm}

\bibitem{AvilaHubSkrip2}
A. Avila, P. Hubert, A. Skripchenko.,
On the Hausdorff dimension of the Rauzy gasket.,
{\it Bulletin de la societe mathematique de France}, 
2016, {\bf 144} (3), pp. 539 - 568.

\vspace{1mm}

\bibitem{DynMalNovUMN} 
I.A. Dynnikov, A.Ya. Maltsev, S.P. Novikov,
Geometry of quasi-periodic functions on the plane,  
{\it Russian Math. Surveys} {\bf 77} : 6, 1061–1085
(2022), arXiv:2306.11257

\vspace{1mm}

\bibitem{DynHubSkrip}
Ivan Dynnikov, Pascal Hubert, Alexandra Skripchenko,
Dynamical Systems Around the Rauzy Gasket and Their Ergodic Properties,
{\it International Mathematics Research Notices IMRN}  
Volume 2023, Issue 8, April 2023, Pages 6461–6503

\vspace{1mm}

\bibitem{NovKvazFunc} 
S.P. Novikov,
Levels of quasiperiodic functions on a plane, and Hamiltonian systems,  
{\it Russian Math. Surveys}, {\bf 54} (5) (1999), 1031-1032

\vspace{1mm}

\bibitem{DynNov} 
I.A. Dynnikov, S.P. Novikov,
Topology of quasi-periodic functions on the plane,  
{\it Russian Math. Surveys}, {\bf 60} (1) (2005), 1-26

\vspace{1mm}

\bibitem{BigQuas} 
A.Ya. Maltsev, 
On the Novikov problem with a large number of quasiperiods and its 
generalizations, Proc. Steklov Inst. Math., 325, 163-176 (2024)

\vspace{1mm}

\bibitem{Sirius} 
A.Ya. Maltsev, 
The Novikov problem for dihedral symmetry potentials, 
J. Math. Sci. (New York) 294(6), 906-914 (2025)

\vspace{1mm}

\bibitem{AnnPhys}
A.Ya. Maltsev, S.P. Novikov, 
Open level lines of a superposition of periodic potentials on a plane, 
{\it Annals of Physics} {\bf 447}(Pt.2), 169039 (2022)

\vspace{1mm}

\bibitem{Superpos} A.Ya. Maltsev, 
On the Novikov Problem for Superposition of Periodic Potentials, 
{\it Regul. Chaot. Dyn.} {\bf 31}, 632-667 (2026).

\vspace{1mm}

\bibitem{SymSuperPos} A.Ya. Maltsev, 
On the level lines of two-layer symmetric potentials, 
arXiv:2025arXiv250115867M 

\vspace{1mm}

\bibitem{Stauffer} 
D. Stauffer, Scaling theory of percolation clusters,
{\it Physics Reports}, Volume 54, Issue 1 (1979), 1-74.

\vspace{1mm}

\bibitem{Essam} 
J.W. Essam, Percolation theory,
{\it Rep. Prog. Phys.} {\bf 43} (1980), 833-912

\vspace{1mm}

\bibitem{Riedel} Eberhard K. Riedel, 
The potts and cubic models in two dimensions: 
A renormalization-group description,
{\it Physica A: Statistical Mechanics and its Applications.,}
Volume 106, Issues 1-2 (1981), 110-121.

\vspace{1mm}

\bibitem{Trugman} S.A. Trugman,
Localization, percolation, and the quantum Hall effect,
{\it Phys. Rev. B} {\bf 27} (1983), 7539-7546

\vspace{1mm}

\bibitem{Arnold1988}
V.I. Arnold,
Remarks on quasicrystallic symmetries,
{\it Physica D} {\bf 33}, 21-25 (1988)



\end{thebibliography}
\end{document}